# Modulation instability induced by cross-phase modulation in a dual-wavelength dispersion-managed soliton fiber ring laser

Z.-C. Luo,* A.-P. Luo, W.-C. Xu, J.-R. Liu, and H.-S. Yin

*Key Laboratory of Photonic Information Technology of Guangdong Higher Education Institutes, School of Information and Optoelectronic Science and Engineering, South China Normal University, Guangzhou, Guangdong 510006, P.R. China*

*Corresponding author: zcluo@scnu.edu.cn

**Abstract**: We report on the observation of modulation instability induced by cross-phase modulation in a dual-wavelength operation dispersion-managed soliton fiber ring laser with net negative cavity dispersion. The passively mode-locked operation is achieved by using nonlinear polarization rotation technique. A new type of dual-wavelength operation, where one is femtosecond pulse and the other is picosecond pulse operation, is obtained by properly rotating the polarization controllers. When the dual-wavelength pulses are simultaneously circulating in the laser ring cavity, a series of stable modulation sidebands appears in the picosecond pulse spectrum at longer wavelength with lower peak power due to modulation instability induced by cross-phase modulation between the two lasing wavelengths. Moreover, the intensities and wavelength shifts of the modulation sidebands can be tuned by varying the power of the femtosecond pulse or the lasing central wavelengths of the dual-wavelength pulses. The theoretical analysis of the modulation instability induced by cross-phase modulation in our fiber laser is also presented.

# 1. Introduction

Modulation instability (MI) as a well-known nonlinear phenomenon has attracted considerable interest [1-9]. MI can be found in many fields of physics, such as nonlinear optics, fluid dynamics and plasma physics. This phenomenon is characterized by an exponential growth of weak optical perturbations from the steady state. Ostrovskii first predicted the presence of MI in nonlinear optics [10]. Hasegawa et al. originally studied the MI in optical fibers [11] and their theoretical prediction was observed experimentally by Tai et al. [12]. When two or more optical fields with different frequencies co-propagate in a fiber, they can interplay through the cross-phase modulation (XPM) mechanism. The interaction of two optical fields through XPM causes an optical field from a stable state to an unstable state, as we called XPM-induced MI. XPM-induced MI can occur between the two polarization components of a single beam or two beams with different wavelengths which co-propagate in a fiber. In the experiments, the pump-probe configurations are most commonly used to observe XPM-induced MI where one of the beams propagates in the normal dispersion region while the other beam experiences anomalous dispersion. Recently, XPM-induced MI in the advanced nonlinear materials, such as the photonics crystal fibers (PCFs) and silicon photonic wires, attracted much attention due to the specially designed characteristics of the materials [13-17].

MI also exists in the soliton fiber laser systems. Although the Kelly sideband was deemed as a result of MI [18], it was later found that Kelly sideband generation was caused by the constructive interference between soliton pulse and dispersive waves [19-20]. The spectral effects caused by MI in the soliton fiber ring lasers have been reported elsewhere, which were characterized by subsideband generation. It was shown that MI with subsideband generation in the fiber ring lasers could be caused by the periodic dispersion variation [21] and periodic power fluctuation [22]. The MI discussed above occurs in the fiber lasers with only single wavelength operation.

When one refers to the dual-wavelength operation fiber lasers, the two lasing wavelengths could experience the XPM interaction between them in the laser cavity. However, there are no evident spectral differences observed in the previous reports because the two lasing wavelengths were the CW or picosecond pulses in the dual-wavelength fiber lasers, it lacked the strong pump beam in the laser cavity as the condition of XPM-induced MI [23-26]. In this paper, we report on the observation of XPM-induced MI between two different lasing wavelengths in a dual-wavelength operation dispersion-managed soliton fiber ring laser with net negative cavity dispersion. A new type of dual-wavelength operation, one is femtosecond pulse and the other is picosecond pulse operation, is obtained by properly adjusting the orientations of the polarization controllers (PCs). When the two pulses simultaneously circulate in the laser ring cavity, they interact through the XPM mechanism. Meanwhile, a series of stable modulation sidebands appears in the optical spectrum of the picosecond pulse at longer wavelength with lower peak power because of XPM-induced MI. In addition, the intensities and wavelength shifts of the modulation sidebands can be tuned by varying the intensity of the femtosecond pulse or the lasing central wavelength locations of the dual-wavelength pulses. The theoretical analysis of XPM-induced MI in our fiber laser is also presented.

## 2. Experimental setup and results

The schematic of the all-fiber ring laser is shown in Fig. 1. The fiber laser has a ring cavity of 19.7 m long with a fundamental repetition 10.23 MHz. A 5 m long erbium-doped fiber (EDF) with dispersion parameter $D \approx -15 \, ps/(nm \cdot km)$ and a 14.7 m long single-mode fiber (SMF) with $D = 17 \, ps/(nm \cdot km)$ comprise the ring cavity. A 980 nm /1550 nm wavelength division multiplexer (WDM) is used to launch the 980 nm pumping laser diode into the laser ring cavity. Two PCs are employed to adjust the polarization states of the pulses in the cavity. Unidirectional operation and polarization selectivity are provided by the polarization-dependent isolator (PD-ISO). The output is taken by a 10 % fiber wideband coupler. In order to prevent any movement and any change of fiber birefringence, all the fibers were fastened to the optical table. An optical spectrum analyzer (OSA) and an oscillograph behind the output coupler are used to study the soliton pulse spectrum and the output pulse-train, respectively.

The nonlinear polarization rotation (NPR) technique was used for achieving the self-started mode-locking state of the fiber laser. The mode-locking threshold was about 30 mW in our fiber laser. In the experiment, when we increased the 980 nm pump power above the mode-locking threshold, the fiber laser achieved the mode-locking state by properly adjusting the PCs. Keep the orientations of the PCs, the self-started mode-locking operation could be attained again by simply increasing the pump power. Depending on the settings of the cavity parameters, such as the orientations of the PCs and the 980 nm pump power, a new type of dual-wavelength operation, one is femtosecond pulse and the other is picosecond pulse operation, was obtained. Fig. 2 (a) shows a typical spectrum of the dual-wavelength lasing operation of the fiber laser. The two lasing wavelengths centered at ~1562 nm and ~1605 nm. The spectrum of the femtosecond pulse at ~1562 nm exhibited sharp, discrete spectral sidebands, as we called Kelly sideband, which were related to the nature of soliton pulses. The Kelly spectral sidebands were well understood as a result of the constructive interference between the soliton pulse and the dispersive waves. When the dual-wavelength operation was achieved, more than one pair of modulation sidebands of the picosecond pulse at longer wavelength was obtained in our experiment, which was similar to the results in Ref. [27]. As can be seen in Fig. 2, there were 4 pairs of modulation sidebands. It is shown in the following theoretical analysis section that the multiple pairs of modulation sidebands are the characteristics of XPM-induced MI in the dispersion-managed fiber link with net negative dispersion. For better clarity of the modulation sidebands, we have shown the enlargement of the portion at ~1605 nm in Fig. 2 (b). It is clearly seen in Fig. 2 (b) that a series of stable modulation sidebands appears in the spectrum due to XPM-induced MI between the two lasing wavelengths. The intensity of the modulation sidebands exhibited a little asymmetry. However, the modulation sidebands located symmetrically at the two sides of the central peak of the picosecond pulse spectrum.

In the experiment, it is worthy to note that the switching operation of single wavelength lasing either at ~1560 nm or at ~1605 nm could be achieved by simply adjusting the orientations of the PCs. It is because the transmission function of the fiber ring laser using the NPR technique can be tuned when the PCs are rotated [28-30]. The typical pulse-trains of the single wavelength lasing operation of femtosecond and picosecond pulse are shown in Fig. 3 (a) and (b), respectively. The intensity of the pulse in Fig. 3 (a) was much higher than that of the pulse shown in Fig. 3 (b). Therefore, we can consider the femtosecond pulse at shorter wavelength with higher peak power as the pump light and the picosecond pulse at longer wavelength with lower peak power as the probe light. In the experimental observation, as long as the dual-wavelength operation was obtained, the femtosecond pulse was always obtained at shorter wavelength region while the picosecond pulse was at longer wavelength region. In the presence of the femtosecond pump pulse, a visible change of the probe pulse spectrum was the appearance of a series of stable modulation sidebands. The generation of the modulation sidebands indicates that the XPM-induced MI occurs between the two lasing wavelengths.

One of the observed aspects of XPM-induced MI is the dependence of the intensity of the modulation sidebands on the pump power. In the experiment, by carefully rotating the PCs, we could decrease the pump power of the femtosecond pulse continuously. Nevertheless, in this case the intensity of the probe light could almost keep invariable. Fig. 4 shows the experimental results for the effect of the pump power on XPM-induced MI sidebands by only slightly rotating one PC while the other cavity parameters were fixed. Fig. 4 (a), (c), (e), (g) present the dual-wavelength operation with a large measured wavelength range from 1540 nm to 1610 nm. Fig. 4 (b), (d), (f), (h) show the evolution of the modulation sidebands which are the enlargement portions of the probe pulse spectra corresponding to Fig. 4 (a), (c), (e), (g). Specifically, in Fig. 4 (a), the average output power from the 10 % coupling port was ~3.8 mW for the femtosecond pulse and ~0.8 mW for the picosecond pulse. Therefore, we can deduce that the average optical power is 38 mW in the laser cavity for the femtosecond pulse and 8 mW for the picosecond pulse. In addition, the 3 dB bandwidth of the femtosecond pump pulse is 5.12 nm in Fig. 4 (a). Correspondingly, the pulse duration, assumed as the fit of hyperbolic secant pulse shape, is ~550 fs. The average pump power of the femtosecond pulse at ~1560 nm circulating in the ring cavity is 28, 16, 0 mW in Fig. 4 (c), (e), (g), respectively. As can be seen in Fig. 4, the number of the modulation sidebands caused by XPM-induced MI changed from 4 to 0 pairs. Initially, there are 4 pairs of the modulation sidebands which have strong spectral strength. When we rotated the PCs, the intensities of the modulation sidebands decreased continuously due to the decreasing power of the femtosecond pulse. Note that the femtosecond pulse no longer existed in the laser cavity in Fig. 4 (g). Consequently, the effect of XPM-induced MI vanished and the modulation sidebands disappeared. It is also to note that the gain bandwidth of MI and the wavelength shifts of the modulation sidebands are related to the power levels, dispersion relation and polarization states of two co-propagating beams. Therefore, the gain bandwidth and the wavelength shifts varied slightly in Fig. 4. However, the general tendencies in the behavior of the output spectra of the probe pulse as a function of the pump power are well consistent with the prediction of the standard theory of XPM-induced MI.

When the orientations of the PCs were further adjusted, we could obtain different lasing central wavelengths of the dual-wavelength operation. Fig. 5 shows the spectrum of the dual-wavelength operation with the pump pulse at 1554.05 nm and the probe pulse at 1586.72 nm. As shown in Fig. 5, comparing to the pump and probe pulses at 1559.77 nm and 1601.56 nm respectively in Fig. 4 (a), the modulation sidebands with larger wavelength shifts and narrower gain bandwidths were observed due to the variation of the lasing central wavelengths and polarization states between the dual-wavelength pulses. For the purpose of better comparison, Fig. 6 illustrates the enlargement of the modulation sidebands of the probe pulses spectra corresponding to Fig. 4 (a) (red curve) and Fig. 5 (green curve). In the case of the probe light lasing at 1586.72 nm, the separation of the first order modulation sideband from the central wavelength of probe pulse was 1.71 nm (1.29 nm for the red curve) and the 3 dB bandwidth of the modulation sideband is 0.14 nm (0.43 nm for the red curve).

In addition, we were able to obtain a different spectrum type of the probe light caused by XPM-induced MI through properly changing the orientations of the PCs. These spectra were characterized by a dip in the central wavelength of the probe pulse spectrum, which separated the probe pulse spectrum into two peaks, as shown in Fig. 7. It is because that the gain spectrum of the MI changes as we rotate the PCs. The intensities of the two peaks besides the central wavelength as well as the wavelength separation from the central wavelength of the probe pulse are equal. In addition, we could tune the dip depth by slightly adjusting the PCs. However, as mentioned above, once the femtosecond pump pulse no longer existed in the laser cavity, the effects of XPM-induced MI on the spectrum of the probe pulse vanished and the modulation sidebands disappeared. As a result, the dual-peak of the probe pulse incorporated into a single peak, which was similar with the spectrum shown in Fig. 4 (h).

## 3. Theoretical analysis of XPM-induced MI

Our fiber ring laser cavity is composed of two characteristic fibers with different dispersion profiles, where one is SMF with negative dispersion in 1550 nm waveband and the other is EDF with normal dispersion. Since the laser has a ring structure, it can be equivalent to a periodic dispersion-managed fiber system for the propagating pulses with a piece of SMF and a piece of EDF in one period. As we concentrate on the XPM-induced MI in dispersion-managed transmission system, therefore, we introduce a model of two pulses co-propagating in a dispersion-managed fiber link to analyze the XPM-induced MI in our fiber ring laser. The analysis starts from the nonlinear Schrodinger equations with two pulses co-propagating in optical fibers [1]:

$$\frac{\partial A_1}{\partial z} = -\frac{1}{v_{g1}(z)}\frac{\partial A_1}{\partial t} - \frac{i}{2}\beta_{21}(z)\frac{\partial^2 A_1}{\partial t^2} - \frac{\alpha(z)}{2}A_1 + i\gamma(z)(|A_1|^2 + 2|A_2|^2)A_1 \tag{1a}$$

$$\frac{\partial A_2}{\partial z} = -\frac{1}{v_{g2}(z)}\frac{\partial A_2}{\partial t} - \frac{i}{2}\beta_{22}(z)\frac{\partial^2 A_2}{\partial t^2} - \frac{\alpha(z)}{2}A_2 + i\gamma(z)(|A_2|^2 + 2|A_1|^2)A_2 \tag{1b}$$

where $A_j(j=1,2)$ is the slowly varying amplitude of the propagating pulse, $v_{gj}$ is the group velocity, $\beta_{2j}$ is the group velocity dispersion (GVD) parameter, $\alpha$ and $\gamma$ are the loss coefficient and the Kerr coefficient, respectively. Here, in order to simplify the analysis and emphasize the effect of the dispersion, the loss and Kerr coefficients are supposed to be the same value for two pulses with different wavelengths in each fiber segment, respectively. In the periodic dispersion-managed system, $\beta_{2j}$ can be expressed as:

$$\beta_{21} = \begin{cases} \beta_{211} < 0, & nL_{map} < z \leq nL_{map} + L_1 \\ \beta_{212} > 0, & nL_{map} + L_1 < z \leq (n+1)L_{map} \end{cases}$$

$$\beta_{22} = \begin{cases} \beta_{221} < 0, & nL_{map} < z \leq nL_{map} + L_1 \\ \beta_{222} > 0, & nL_{map} + L_1 < z \leq (n+1)L_{map} \end{cases} \quad (2)$$

where $n = 0, 1, 2...$, $L_{map}$ is the period of the dispersion management (in our fiber ring laser, it is equal to the total length of SMF and EDF). $L_1$ is the length of the first fiber segment in one period. Then we assume that the propagating pulse in each period has the form:

$$A_j(z,t) = a_j(z,t) \exp(-\int_0^z \alpha(z')/2 dz') \quad (3)$$

Substituting Eq. (3) into Eq. (1), we obtain:

$$\frac{\partial a_1}{\partial z} = -\frac{1}{v_{g1}(z)} \frac{\partial a_1}{\partial t} - \frac{i}{2} \beta_{21}(z) \frac{\partial^2 a_1}{\partial t^2} + if(z)(|a_1|^2 + 2|a_2|^2) a_1 \quad (4a)$$

$$\frac{\partial a_2}{\partial z} = -\frac{1}{v_{g2}(z)} \frac{\partial a_2}{\partial t} - \frac{i}{2} \beta_{22}(z) \frac{\partial^2 a_2}{\partial t^2} + if(z)(|a_2|^2 + 2|a_1|^2) a_2 \quad (4b)$$

where $f(z) = \gamma(z) \exp(-\int_0^z \alpha(z') dz')$ is the periodic function with the same period of dispersion management fiber link. The steady solution of Eq. (4) is:

$$a_j(z,t) = \sqrt{P_j} \exp[i(P_j + 2P_{3-j}) \int_0^z f(z') dz'] \quad (5)$$

where $P_j$ denotes the power of the input pulse. Then we perturb Eq. (4) by a small amplitude fluctuation $b_j(z,t)$ added to Eq. (5):

$$a_j(z,t) = (\sqrt{P_j} + b_j) \exp[i(P_j + 2P_{3-j}) \int_0^z f(z') dz'] \quad (6)$$

Inserting Eq. (6) into Eq. (4) and using $T = t - z/v_{g1}$, one set of two equations is obtained for the

fluctuation:

$$\frac{\partial b_1}{\partial z} = -\frac{i}{2}\beta_{21}(z)\frac{\partial^2 b_1}{\partial T^2} + if(z)[P_1(b_1 + b_1^*) + 2\sqrt{P_1 P_2}(b_2 + b_2^*)] \tag{7a}$$

$$\frac{\partial b_2}{\partial z} = -\delta(z)\frac{\partial b_2}{\partial T} - \frac{i}{2}\beta_{22}(z)\frac{\partial^2 b_2}{\partial T^2} + if(z)[P_2(b_2 + b_2^*) + 2\sqrt{P_1 P_2}(b_1 + b_1^*)] \tag{7b}$$

where $\delta(z) = 1/v_{g2} - 1/v_{g1}$ is the walk-off parameter, the superscript $*$ indicates the counterpart complex conjugate. Since $f(z)$, $\beta_{2j}(z)$, and $\delta(z)$ are periodic functions of the dispersion management system, they can be expanded as complex Fourier series with fundamental wave constant $k$ ($k = 2\pi / L_{map}$):

$$f(z) = \sum_{n=-\infty}^{\infty} F_n \exp(iknz), \quad \beta_{21}(z) = \sum_{n=-\infty}^{\infty} G_n \exp(iknz),$$

$$\beta_{22}(z) = \sum_{n=-\infty}^{\infty} H_n \exp(iknz), \quad \delta(z) = \sum_{n=-\infty}^{\infty} \sigma_n \exp(iknz) \tag{8}$$

where $F_n$, $G_n$, $H_n$ and $\sigma_n$ denote the Fournier series coefficients of $f(z)$, $\beta_{21}(z)$, $\beta_{22}(z)$ and $\delta(z)$, respectively.

Because the dispersion-managed system could be treated as a grating, we supposed that the fluctuation with wave constant $k_p$ satisfies the phase-matched condition. Then we introduce the following transformation of $b_j(z)$:

$$b_j = c_j \exp(-ik_p z / 2), \tag{9}$$

where $k_p = 2p\pi / L_{map}$. Substituting Eq. (8) and (9) back to Eq. (7), the following equations can be obtained:

$$\frac{\partial c_1}{\partial z} = -\frac{i}{2}G_0\frac{\partial^2 c_1}{\partial T^2} + \frac{i}{2}k_p c_1 + iF_0 P_1 c_1 + iF_p^* P_1 c_1^* + 2i\sqrt{P_1 P_2}F_0 c_2 + 2i\sqrt{P_1 P_2}F_p^* c_2^* \tag{10a}$$

$$\frac{\partial c_2}{\partial z} = -\sigma_0 \frac{\partial c_2}{\partial T} - \frac{i}{2} H_0 \frac{\partial^2 c_2}{\partial T^2} + \frac{i}{2} k_p c_2 + i F_0 P_2 c_2 + i F_p^* P_2 c_2^* + 2i\sqrt{P_1 P_2} F_0 c_1 + 2i\sqrt{P_1 P_2} F_p^* c_1^* \qquad (10b)$$

Then we made the Fourier transform of Eq. (10) to get the following equations:

$$\frac{\partial \hat{c}_1}{\partial z} = \frac{i}{2} G_0 \omega^2 \hat{c}_1 + \frac{i}{2} k_p \hat{c}_1 + i F_0 P_1 \hat{c}_1 + i F_p^* P_1 \hat{c}_1^* + 2i\sqrt{P_1 P_2} F_0 \hat{c}_2 + 2i\sqrt{P_1 P_2} F_p^* \hat{c}_2^* \qquad (11a)$$

$$\frac{\partial \hat{c}_2}{\partial z} = -i\sigma_0 \omega \hat{c}_2 + \frac{i}{2} H_0 \omega^2 \hat{c}_2 + \frac{i}{2} k_p \hat{c}_2 + i F_0 P_2 \hat{c}_2 + i F_p^* P_2 \hat{c}_2^* + 2i\sqrt{P_1 P_2} F_0 \hat{c}_1 + 2i\sqrt{P_1 P_2} F_p^* \hat{c}_1^* \qquad (11b)$$

Next, defining the transformation:

$$\hat{c}_j = R_j + iI_j, \quad F_p = F_{1p} + iF_{2p} \qquad (12)$$

By introducing the transformation shown in Eq. (12), Eq. (11) can be written as:

$$\frac{\partial}{\partial z}\begin{pmatrix} R_1 \\ I_1 \\ R_2 \\ I_2 \end{pmatrix} = \begin{bmatrix} F_{2p}P_1 & -D_1 + F_{1p}P_1 & 2\sqrt{P_1P_2}F_{2p} & 2\sqrt{P_1P_2}(F_{1p}-F_0) \\ D_1 + F_{1p}P_1 & -F_{2p}P_1 & 2\sqrt{P_1P_2}(F_{1p}+F_0) & -2\sqrt{P_1P_2}F_{2p} \\ 2\sqrt{P_1P_2}F_{2p} & 2\sqrt{P_1P_2}(F_{1p}-F_0) & F_{2p}P_2 & -D_2 + F_{1p}P_2 \\ 2\sqrt{P_1P_2}(F_{1p}+F_0) & -2\sqrt{P_1P_2}F_{2p} & D_2 + F_{1p}P_2 & -F_{2p}P_2 \end{bmatrix} \begin{pmatrix} R_1 \\ I_1 \\ R_2 \\ I_2 \end{pmatrix} \qquad (13)$$

where $D_1 = \frac{1}{2} G_0 \omega^2 + \frac{1}{2} k_p + F_0 P_1$, $D_2 = \frac{1}{2} H_0 \omega^2 + \frac{1}{2} k_p + F_0 P_2 - \sigma_0 \omega$.

From the eigenvalues of Eq. (13), we can obtain the power gain $\lambda(\omega)$ for the p-th modulation sideband caused by XPM-induced MI effect:

$$\lambda(\omega) = \operatorname{Re}(\frac{1}{2}\sqrt{d_1 + 2\sqrt{d_2}}) \qquad (14)$$

where $d_1 = 2(|F_p|^2 - F_0^2)(P_1^2 + 8P_1P_2 + P_2^2) - 2(e_1^2 + e_2^2) - 4F_0(e_1P_1 + e_2P_2)$

$$d_2 = (|F_p|^2 - F_0^2)^2 (P_1 + P_2)^2 (P_1^2 + 14 P_1 P_2 + P_2^2) + F_0^2 (e_1 + e_2)^2$$
$$\times (P_1^2 + 14 P_1 P_2 + P_2^2) + F_0^2 [4(e_1^2 - e_2^2)(P_1^2 - P_2^2)$$
$$+ (e_1 - e_2)^2 (P_1 + P_2)^2] + 2|F_p|^2 \times [(e_2^2 - e_1^2)(P_1^2 - P_2^2)$$
$$- 8(e_1 - e_2)^2 P_1 P_2] + 4 F_0 |F_p|^2 [e_1 (P_1^3 + 7 P_1 P_2^2 + 8 P_1^2 P_2)$$
$$+ e_2 (P_2^3 + 7 P_1^2 P_2 + 8 P_1 P_2^2)] + 2(e_1 + e_2) F_0^3 [P_1^3 + 15 P_1^2 P_2 + 15 P_1 P_2^2 + P_2^3]$$
$$+ 2(e_1 - e_2) F_0^3 (P_1 - P_2)(P_1 + P_2)^2$$
$$+ (e_1^2 - e_2^2)^2 + 4 F_0 (e_1^2 - e_2^2)(e_1 P_1 - e_2 P_2)$$

and $e_1 = \frac{1}{2} G_0 \omega^2 + \frac{k_p}{2}$, $e_2 = \frac{1}{2} H_0 \omega^2 + \frac{k_p}{2} - \sigma_0 \omega$, $F_0 = \frac{1}{L_{map}} [\int_0^{L_1} \gamma_1 e^{-\alpha_1 z} dz + \int_{L_1}^{L_{map}} \gamma_2 e^{-\alpha_2 z} dz]$,

$F_p = \frac{1}{L_{map}} [\int_0^{L_1} \gamma_1 e^{-\alpha_1 z} e^{-i 2 p \pi z / L_{map}} dz + \int_{L_1}^{L_{map}} \gamma_2 e^{-\alpha_2 z} e^{-i 2 p \pi z / L_{map}} dz]$, $G_0 = \frac{1}{L_{map}} [\int_0^{L_1} \beta_{211} dz + \int_{L_1}^{L_{map}} \beta_{212} dz]$,

$H_0 = \frac{1}{L_{map}} [\int_0^{L_1} \beta_{221} dz + \int_{L_1}^{L_{map}} \beta_{222} dz]$, $\sigma_0 = \frac{1}{L_{map}} [\int_0^{L_1} \delta_1 dz + \int_{L_1}^{L_{map}} \delta_2 dz]$.

Here, $\gamma_j$ and $\alpha_j$ are the Kerr coefficient and the loss coefficient of the two segment fibers with different dispersion profiles in a dispersion-managed period, respectively. $\delta_j$ is the walk-off parameter in two segment fibers.

Now we analyze the power gain $\lambda(\omega)$ for the p-th modulation sideband caused by XPM-induced MI effect according to Eq. (14), two cases are considered in the following calculations:

1) The two signals have the same or nearly equivalent dispersion coefficients in each segment fiber and the walk-off effect can be neglected (for example, the adjacent channels in WDM system), then $d_j$ could be simplified to:

$$d_1 = 2(|F_p|^2 - F_0^2)(P_1^2 + 8 P_1 P_2 + P_2^2) - 4(\frac{G_0}{2} \omega^2 + \frac{k_p}{2})^2 - 4 F_0 (P_1 + P_2)(\frac{G_0}{2} \omega^2 + \frac{k_p}{2}),$$

$$d_2 = (|F_p|^2 - F_0^2)^2 (P_1 + P_2)^2 (P_1^2 + 14P_1P_2 + P_2^2) - 4F_0(|F_p|^2 - F_0^2)(\frac{G_0}{2}\omega^2 + \frac{k_p}{2}) \\ \times [P_1^3 + 15P_1^2P_2 + 15P_1P_2^2 + P_2^3] + 4F_0^2(\frac{G_0}{2}\omega^2 + \frac{k_p}{2})^2(P_1^2 + 14P_1P_2 + P_2^2) \quad (15)$$

Fig. 8 shows the gain spectra of XPM-induced MI calculated according to Eq. (14) and (15) with optical powers $P_1 = 10mW$, $P_2 = 8mW$ and the nonlinear coefficients $\gamma_1 = 3W^{-1}km^{-1}$, $\gamma_2 = 5W^{-1}km^{-1}$. The lengths of the two segment fibers in one period of the dispersion-managed fiber system are $L_1 = 14.7m$ with negative dispersion $\beta_{21} = -23.5 ps^2/km$ and $L_2 = 5m$ with positive dispersion $\beta_{22} = 24 ps^2/km$. Therefore, the net dispersion in one periodic fiber link is negative. As can be seen from Fig. 8, we know that those fluctuations satisfying the grating phase-matched condition could obtain larger gain and contribute to the discrete gain spectra. It is to note that there are multiple pairs of modulation sidebands and the intensities of the sidebands are different from each other. The gain for the higher order sideband is obviously weaker. Then we investigated the influence of the pump power on the gain spectra. Fig. 9 shows the gain spectra with the same calculated parameters as those in Fig. 8 except that the pump power was changed to 40 mW. It is obvious that the gain for each modulation sideband is stronger with larger pump power than that shown in Fig. 8.

2) The two signals in the system have larger wavelength separation. For example, in our laser system, one is 1560 nm, the other is 1600 nm. And then the dispersion difference should be taken into account. Fig. 10 shows the analytical results according to Eq. (14) with $\beta_{211} = -22 ps^2/km$, $\beta_{212} = 19 ps^2/km$, $\beta_{221} = -25 ps^2/km$, $\beta_{222} = 23 ps^2/km$, the other parameters are the same with Fig. 9. Note that the fiber laser has a ring cavity and both pulses circulate in the laser cavity all the time. Therefore, to simplify the calculation, here we also neglect the walk-off effect between the two co-propagating pulses. As can be seen in Fig. 10, the positions and the intensity of the modulation sidebands changed slightly due to the variation of the dispersion relation. However, there still exist multiple pairs of modulation sidebands in the gain spectrum of MI.

## 4. Discussion

In the theoretical calculation, we found that the generation of the multiple modulation sidebands caused by XPM-induced MI could only exist in the dispersion-managed fiber link with net negative dispersion. When the dispersion-managed fiber system has a net positive dispersion, only one pair of the fundamental modulation sidebands (0-th modulation sideband) can be obtained. Note that the net dispersion of our fiber ring laser system is negative, therefore, multiple pairs of modulation sidebands could be obtained in the experiments, which was in agreement with the theoretical analysis. In addition, based on the analytical calculation, we found that the gain peak positions of MI were relevant to the length of the two segment fibers, which was essentially related to the total dispersion value in one period of the dispersion-managed fiber link. By changing the total dispersion value of the dispersion-managed fiber system, one can tune the sideband positions in the calculated gain spectra.

In the experimental observation, the pump pulse spectrum exhibits several peaks which we called Kelly sidebands. However, due to the low power of the Kelly sidebands, their impact on the XPM-induced MI can be neglected. It is to note that two characteristic fibers with negative and normal dispersion were used to construct the fiber ring cavity. Therefore, the effective co-propagation length could cover the whole dispersion-managed map of the laser cavity when the pump and probe pulses interact with each other through XPM mechanism. As the modulation sidebands were excited by XPM-induced MI, like the pump and probe pulses, the modulation sidebands also periodically experience amplification in EDF and loss at the output coupler. And finally the modulation sidebands evolve among these effects to achieve an equilibrium state. Since we did not consider some conditions, such as the periodic gain effect of EDF, the polarization states of two co-propagating pulses and successive circulating round trips of the pulses, into the theoretical model, as a result, the positions and intensities of the modulation sidebands in the experimental observation were not exactly consistent with the calculation results. However, the generally experimental tendencies of the output spectra of the probe pulse are consistent with the theoretical analysis of XPM-induced MI discussed above.

## 5. Conclusion

In conclusion, we have demonstrated the XPM-induced MI in a dual-wavelength dispersion-managed soliton fiber ring laser with net negative dispersion passively mode-locked by using NPR technique. By properly rotating the orientations of the PCs, we obtained a new type of dual-wavelength operation where one is femtosecond pulse and the other is picosecond pulse operation. The two pulses co-propagated in the laser ring cavity and interacted through the XPM mechanism. A series of stable modulation sidebands presented in the spectrum of the picosecond probe pulse due to XPM-induced MI. The intensities and the wavelength shifts of the modulation sidebands can be tuned by changing the power of femtosecond pulse and the lasing central wavelength of the dual-wavelength pulses. In addition, we presented the theoretical analysis of XPM-induced MI in our fiber ring laser.

**Acknowledgement**

The authors wish to thank Prof. Zujie Fang of Shanghai Institute of Optics and Fine Mechanics (CAS, Shanghai) for many useful discussions. This work was supported by the Natural Science Foundation of Guangdong Province, China (Grant No. 04010397), Specialized

Research Fund for the Doctoral Program of Higher Education (Grant No. 20094407110002), and Specialized Research Fund for Innovative Young Scholars of the Higher Education in Guangdong. Z. C. Luo acknowledges the financial support from the Key Program of Scientific Research of South China Normal University, China (Grant No. 09GDKC04).

**References**

[1] G. P. Agrawal, *Nonlinear Fiber Optics*. New York: Academic Press, 1993.

[2] G. P. Agrawal, *Phys. Rev. Lett.*, **59,** 880 (1987).

[3] P. T. Dinda, G. Millot, E. Seve, and M. Haelterman, *Opt. Lett.*, **21**, 1640 (1996).

[4] R. Hui, M. O'Sullivan, A. Robinson, M. Taylor, *J. Lightwave Technol.*, **15**, 1071 (1997).

[5] T. Tanemura and K. Kikuchi, *J. Opt. Soc. Am. B*, **20**, 2502 (2003).

[6] W. C. Xu, S. M. Zhang, W. C. Chen, A. P. Luo, and S. H. Liu, *Opt. Commun*. **199**, 355 (2001).

[7] G. Rossi, D. Amans, E. Brainis, M. Haelterman, Ph. Emplit and S. Massar, *Opt. Lett.*, **30**, 1051 (2005).

[8] P. Kaewplung, T. Angkaew, and K. Kikuchi, *J. Lightwave Technol.*, **20**, 1895 (2002).

[9] X. Dai, Y. Xiang, S. Wen, and D. Fan, *J. Opt. Soc. Am. B*, **26**, 564 (2009).

[10] L. A. Ostrovskii, *Zh. Eksp. Teor. Fiz*. **51**, 1189 (1966). [*Sov. Phys. JETP* **24**, 797 (1967)]

[11] A. Hasegawa and W. F. Brinkman, *IEEE J. Quantum Electron*., **QE-16**, 694 (1980).

[12] K. Tai, A. Hasegawa, and A. Tomita, *Phys. Rev. Lett.*, **56**, 135 (1986).

[13] J. S. Y. Chen, G. K. L. Wong, S. G. Murdoch, R. J. Kruhlak, R. Leonhardt, J. D. Harvey, N. Y. Joly and J. C. Knight, *Opt. Lett.*, **31**, 873 (2006).

[14] A. T. Nguyen, K. P. Huy, E. Brainis, P. Mergo, J. Wojcik, T. Nasilowski, J. V. Erps, H. Thienpont, and S. Massar, *Opt. Express*, **14**, 8290 (2006).

[15] E. E. Serebryannikov, S. O. Konorov, A. A. Ivanov, M. V. Alfimov, M. Scalora, and A. M. Zheltikov, *Phys. Rev. E*, **72**, 027601 (2005).

[16] S. O. Konorov, D. A. Akimov, A. A. Ivanov, M. V. Alfimov, K. V. Dukel'skⅡ, A. V. Khokhlov, V. S. Shevandin, Yu. N. Kondrat'ev, A. M. Zheltikov, *Appl. Phys. B*, **80**, 437 (2005)

[17] I. W. Hsieh, X. Chen, J. I. Dadap, N. C. Panoiu, R. M. Osgood, Jr., S. J. McNab, and Y. A. Vlasov, *Opt. Express*, **15**, 1135 (2007).

[18] D. J. Richardson, R. I. Laming, D. N. Payne, V. I. Matsas, and M. W. Phillips, *Electron. Lett.*, **27**, 1451 (1991).

[19] S. M. J. Kelly, *Electron. Lett.*, **28**, 806 (1992).

[20] Z. C. Luo, A. P. Luo, W. C. Xu, C. X. Song, Y. X. Gao, and W. C. Chen, *Laser Phys. Lett.*, **6**, 582 (2009).

[21] D. Y. Tang, W. S. Man, H. Y. Tam, and M. S. Demokan, *Phys. Rev. A*, **61**, 023804 (2000).

[22] Z. C. Luo, W. C. Xu, C. X. Song, A. P. Luo and W. C. Chen, *Eur. Phys. J. D*., **54**, 693 (2009).

[23] S. Pan and J. P. Yao, *Opt. Express*, **17**, 5414 (2009).

[24] J. B. Schlager, S. Kawanishi and M. Saruwatari, *Electron. Lett.*, **27**, 2072 (1991).

[25] H. Zhang, D. Y. Tang, X. Wu, and L. M. Zhao, *Opt. Express*, **17**, 12692 (2009).

[26] Y. D. Gong, X. L. Tian, M. Tang, P. Shum, M. Y. W. Chia, V. Paulose, J. Wu, and K. Xu, *Opt. Commun.*, **265**, 355 (2001).

[27] A. S. Gouveia-Neto, M. E. Faldon, A. S. B. Sombra, P. G. J. Wigley, and J. R. Taylor, *Opt. Lett.* **13**, 901 (1988).

[28] A. D. Kim, J. N. Kutz, and D. J. Muraki, *IEEE J. Quantum Electron.*, **36**, 465 (2000).

[29] Z. C. Luo, W. C. Xu, C. X. Song, A. P. Luo, and W. C. Chen, *Chin. Phys. B*, **18**, 2328 (2009).

[30] D. Y. Tang, L. M. Zhao, B. Zhao, and A. Q. Liu, *Phys. Rev. A*, **72**, 043816 (2005).

**Figure Captions**

Fig. 1. Schematic of the fiber ring laser cavity. PC: polarization controller; PD-ISO: polarization-dependent isolator; WDM: wavelength division multiplexer.

Fig. 2. (a) A typical spectrum of the dual-wavelength mode-locked operation. (b) The enlargement portion of the picosecond pulse at longer wavelength.

Fig. 3. A typical pulse-train of the single wavelength lasing operation of femtosecond pulse (a) and picosecond pulse (b).

Fig. 4. Observation of the influence of the femtosecond pulse power on the intensities of the modulation sidebands. (a), (c), (e), (g): the dual-wavelength operation with a large measured wavelength range from 1540 nm to 1610 nm. (b), (d), (f), (h): the enlargement portions of the modulation sidebands corresponding to the (a), (c), (e), (g).

Fig. 5. Spectrum of the dual-wavelength operation with different lasing central wavelengths by rotating the PCs.

Fig. 6. Spectra of probe pulses recorded with different PCs settings. The central wavelengths of the two probe beams are 1601.56 nm (red curve) and 1586.72 nm (green curve).

Fig. 7. Dual-wavelength operation with a dip shown in the central wavelength of the probe picosecond pulse spectrum.

Fig. 8. Calculated gain spectra of XPM-induced MI with the parameters of $P_1 = 10mW$, $P_2 = 8mW$, $\beta_{21} = -23.5 ps^2/km$, $\beta_{22} = 24 ps^2/km$, $L_1 = 14.7m$, $L_2 = 5m$, $\gamma_1 = 3W^{-1}km^{-1}$, $\gamma_2 = 5W^{-1}km^{-1}$.

Fig. 9. Calculated gain spectra of XPM-induced MI with same parameters in Fig. 8

except that the pump power was changed to $P_1 = 40mW$.

Fig. 10. Calculated gain spectra incorporated dispersion difference. The GVD coefficients are $\beta_{211} = -22 ps^2/km$, $\beta_{212} = 19 ps^2/km$, $\beta_{221} = -25 ps^2/km$, $\beta_{222} = 23 ps^2/km$. The other parameters are the same with Fig. 9.

Fig. 1.

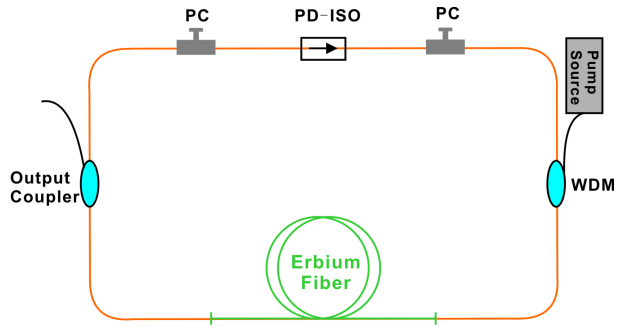

Fig. 2.

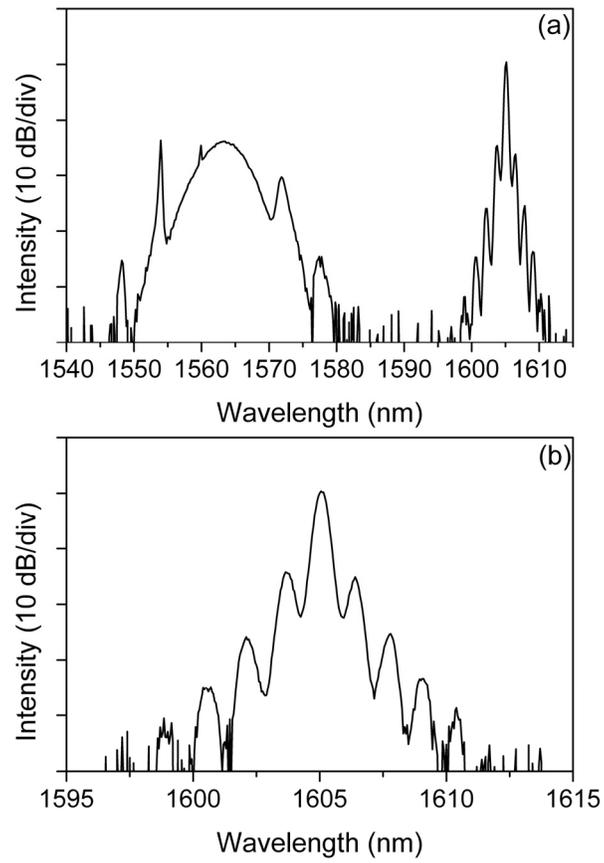

Fig. 3.

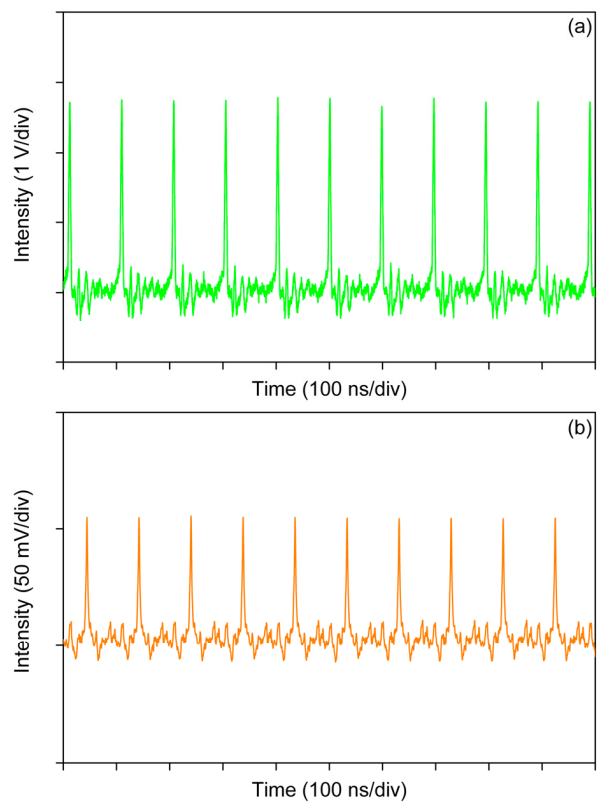

Fig. 4.

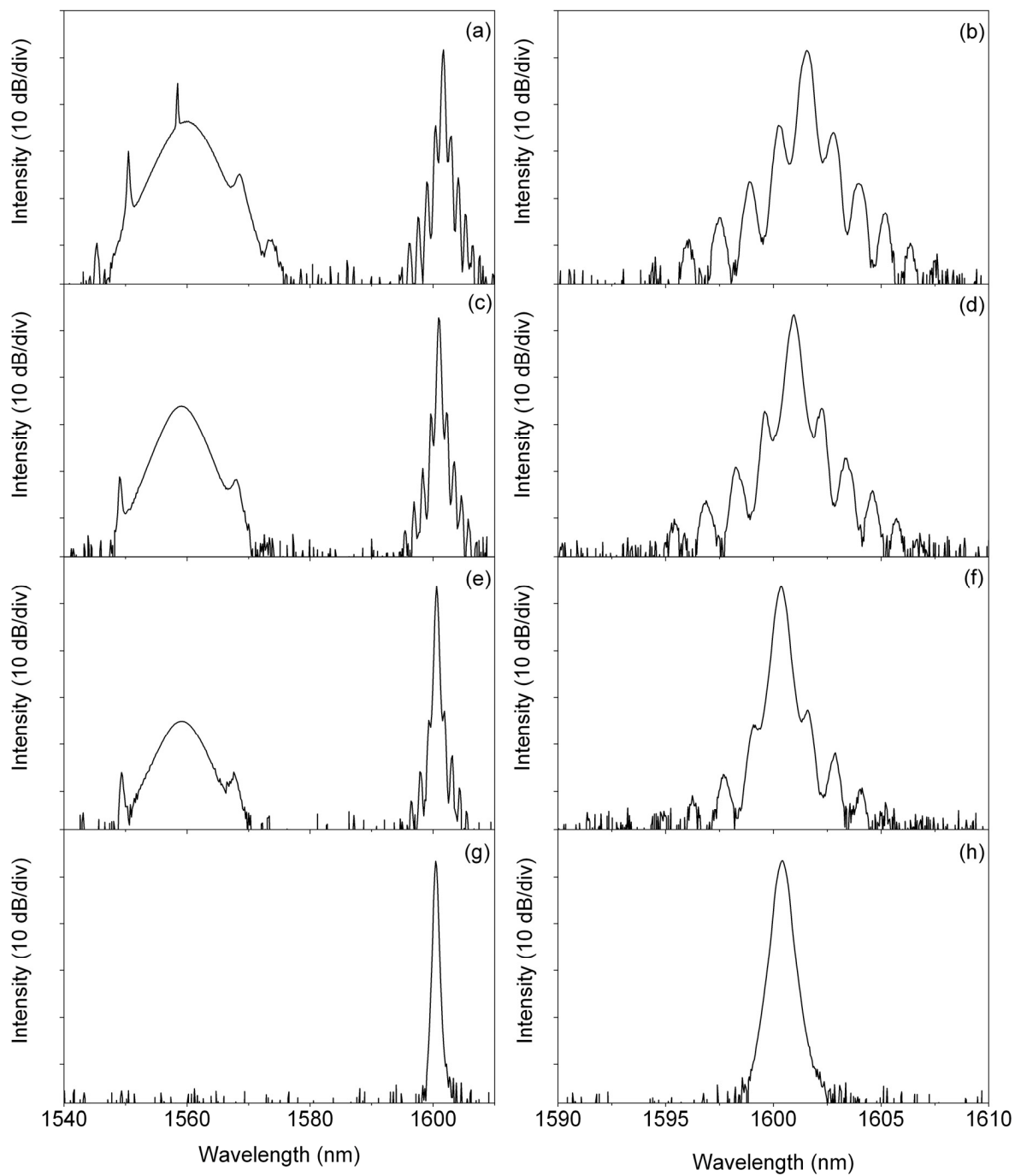

Fig. 5.

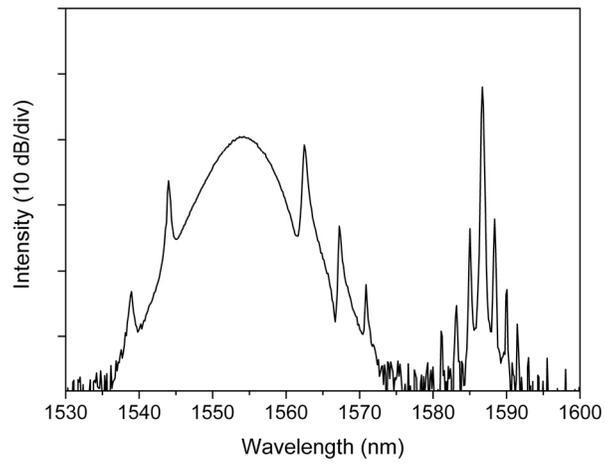

Fig. 6.

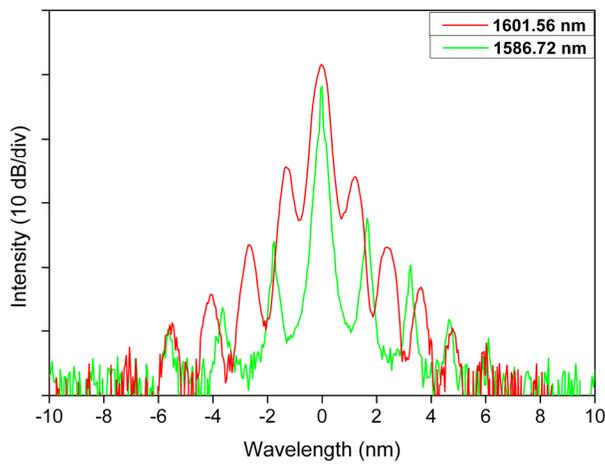

Fig. 7.

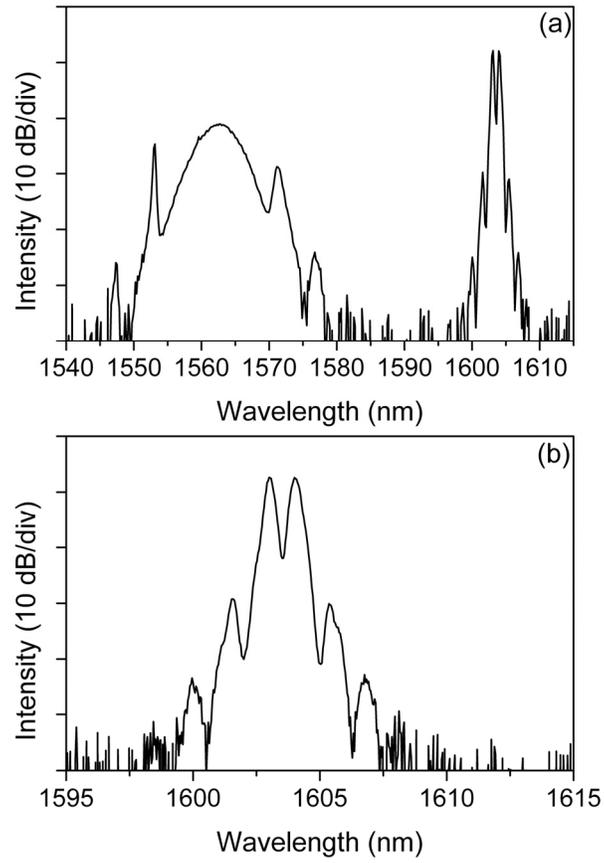

Fig. 8.

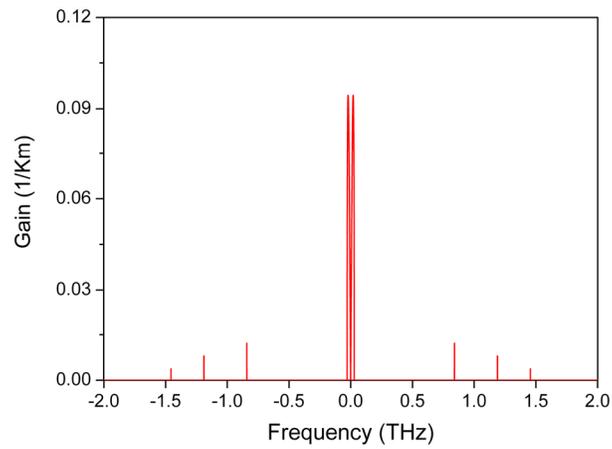

Fig. 9.

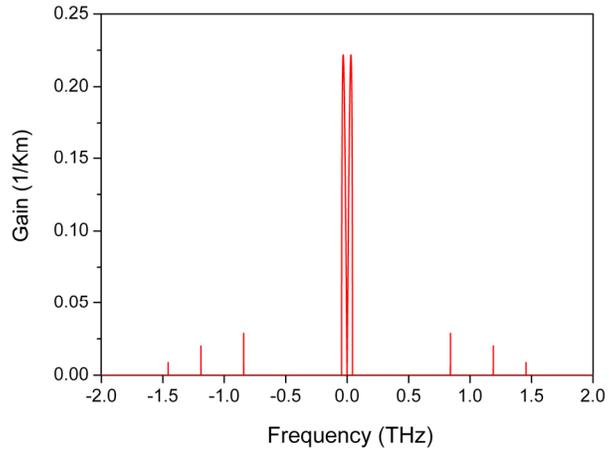

Fig. 10.

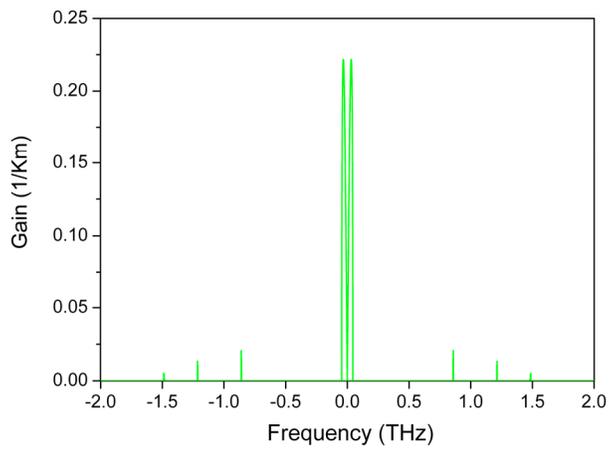